\RequirePackage{fix-cm}
\documentclass[twocolumn,epjc3]{svjour3}  
\smartqed  % flush right qed marks, e.g. at end of proof
\RequirePackage{graphicx}
\usepackage{amsmath}
\usepackage{xspace}
\usepackage{comment}
\usepackage{xcolor}
\usepackage{appendix}
\usepackage{url}
\usepackage{multirow}
\usepackage{tabularx}
\usepackage[export]{adjustbox}
\newcommand{\la}{\langle}
\newcommand{\lla}{\left\langle}
\newcommand{\ra}{\rangle}
\newcommand{\rra}{\right\rangle}
\newcommand{\pt}{\ensuremath{p_\mathrm{T}}\xspace}
\newcommand{\mpt}{\ensuremath{[\pt]}\xspace}
\newcommand{\mmpt}{\ensuremath{\la[\pt]\ra}\xspace}
\newcommand{\mptk}[1]{\ensuremath{[\pt^{(#1)}]}\xspace}
\newcommand{\mmptk}[1]{\ensuremath{\la[\pt^{(#1)}]\ra}\xspace}

\newcommand{\XeXe}{\mbox{Xe--Xe}\xspace}
\newcommand{\pp}{\mbox{pp}\xspace}
\newcommand{\Xe}{\ensuremath{^{129}}Xe\xspace}
\newcommand{\snn}{\ensuremath{\sqrt{s_\mathrm{NN}}}\xspace}
\journalname{Eur. Phys. J. A}
\begin{document}

\title{Generic multi-particle transverse momentum correlations as a new tool for studying nuclear structure at the energy frontier %generic recursive algorithm for multi-particle transverse momentum correlations: a new tool for studying nuclear structure %\thanksref{t1}
}
%\subtitle{Do you have a subtitle?\\ If so, write it here}

\titlerunning{Multi-particle \mpt correlations}        % if too long for running head

\author{Emil Gorm Dahlbæk Nielsen\thanksref{addr1}
        \and
        Frederik K. Rømer\thanksref{addr1} 
        \and 
        Kristjan Gulbrandsen\thanksref{addr1} 
        \and
        You Zhou\thanksref{e1,addr1} 
}

%\thankstext{t1}{Grants or other notes
%about the article that should go on the front page should be
%placed here. General acknowledgments should be placed at the end of the article.
\thankstext{e1}{e-mail: you.zhou@cern.ch}

%\authorrunning{Short form of author list} % if too long for running head

\institute{Niels Bohr Institute, University of Copenhagen, 2200 Copenhagen, Denmark \label{addr1}
}

\date{Received: date / Accepted: date}
% The correct dates will be entered by the editor

\maketitle

\begin{abstract}
The mean transverse momentum of produced particles, \mpt, and its event-by-event fluctuations give direct access to the initial conditions of ultra-relativistic heavy-ion collisions and help probe the colliding nuclei's structure. The \mpt fluctuations can be studied via multi-particle \pt correlations; so far, only the lowest four orders have been studied. Higher-order fluctuations can provide stronger constraints on the initial conditions and improved sensitivity to the detailed nuclear structure; however, their direct implementation can be challenging and is still lacking. In this paper, we apply a generic recursive algorithm for the genuine multi-particle \pt correlations, which enables the accurate study of higher-order \mpt fluctuations without computationally heavy processing for the first time. With this algorithm, we will examine the power of multi-particle \pt correlations through Monte Carlo model studies with different nuclear structures. The impact on the nuclear structure studies, including the nuclear deformation and triaxial structure, will be discussed. These results will demonstrate the usefulness of multi-particle \pt correlations for studying nuclear structure in high-energy nuclei collisions at RHIC and the LHC, which could serve as complementary to existing low-energy nuclear structure studies.
% \PACS{PACS code1 \and PACS code2 \and more}
% \subclass{MSC code1 \and MSC code2 \and more}
\end{abstract}

\section{Introduction}
\label{sec:intro}

The primary goal of the heavy-ion collisions at ultra-relativistic energies, such as at the Relativistic Heavy Ion Collider (RHIC) and the Large Hadron Collider (LHC), is to create a quark-gluon plasma (QGP) and study its properties in the laboratory~\cite{Shuryak:1980tp,Shuryak:1978ij}. Experimental measurements of collective flow~\cite{ALICE:2011ab,ALICE:2016ccg} and its comparison to hydrodynamic model calculations~\cite{Muller:2012zq,Heinz:2013th,Song:2017wtw}, in particular via global Bayesian analysis~\cite{Bernhard:2019bmu,JETSCAPE:2020mzn}, provide a powerful approach to extracting unique information about the specific viscosity of the QGP and its time evolution (temperature dependence). This represents the state-of-the-art understanding of the QGP~\cite{ALICE:2022wpn}. However, the sizable uncertainty of the extracted QGP transport coefficients originates from poor knowledge about the initial conditions of heavy-ion collisions. A further reduction of this uncertainty can be achieved via in-depth investigations of the initial conditions~\cite{Bernhard:2019bmu,JETSCAPE:2020mzn}. Among various collective flow observables, the mean transverse momentum, \mpt, and its event-by-event fluctuations trace their origins to the initial state of the heavy-ion collision~\cite{Broniowski:2009fm,Bozek:2012fw} and are arguably a more direct way to study the initial state fluctuations \cite{Giacalone:2020lbm}. Several initial state properties have been proposed as good estimators of event-by-event mean transverse momentum fluctuations. These properties include the initial size $R$ and the size fluctuations of the system that were introduced in~\cite{Broniowski:2009fm}, the transverse area of the overlap region and the initial eccentricity $\epsilon_2$, introduced as $A_e$ in \cite{Schenke:2020uqq}, as well as the initial energy per unit rapidity at the initial time $\tau_0$, $E_i$ \cite{Gardim:2020sma}. These estimators have been verified using hydrodynamic calculations that show $E_i$ provides a strictly linear correlation to the \mpt \cite{Gardim:2019xjs}. Essentially, the \mpt and its fluctuations reveal information about the fluctuations of the energy deposited in the initial state on an event-by-event basis.

\begin{figure*}[t!]
  \includegraphics[width=\textwidth]{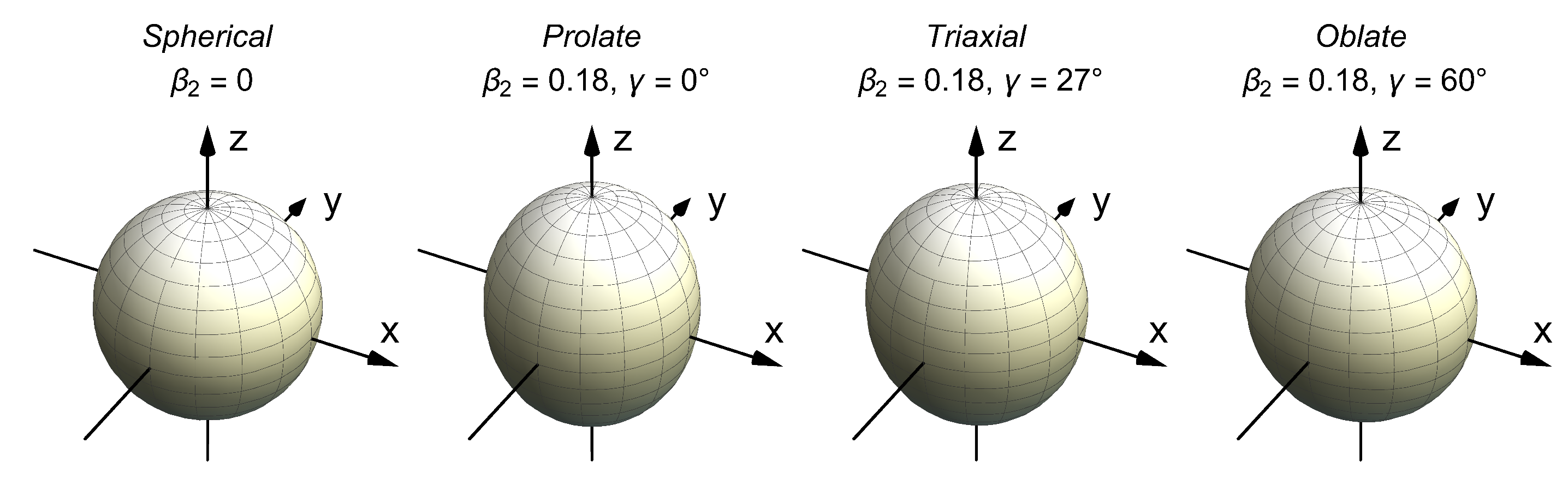}
%  \includegraphics[width=0.22\textwidth]{figs/obj0.pdf}
%  \includegraphics[width=0.22\textwidth]{figs/obj1.pdf}
%  \includegraphics[width=0.22\textwidth]{figs/obj2.pdf}
%  \includegraphics[width=0.22\textwidth]{figs/obj3.pdf}
% figure caption is below the figure
\caption{Illustration of the surface of nuclei with different quadrupole deformation. From left to right: spherical, prolate, triaxial, and oblate. }
\label{fig:0}       % Give a unique label
\end{figure*}

It has been proposed that an improved understanding of the initial conditions of nuclei collisions will allow for a more precise description of the geometrical shape of the overlapping colliding region~\cite{Bally:2022vgo}. In particular, for central collisions where the two colliding nuclei fully overlap, the spatial shape of the overlapping region could directly reflect the nuclear structure that existed before the collision occurred (in the case of body-body collisions). It has also been shown that the imprint of the nuclear structure remains in the subsequent final state of the collective expansion, i.e., \mpt fluctuations~\cite{Bhatta:2021qfk} and anisotropic flow~\cite{Lu:2023fqd}, after the dynamic evolution of nucleus collisions. The study of nuclear structure is an important research topic in nuclear physics, as it reveals the fundamental properties and interactions of nucleons and nuclei. These studies allow us to test and improve the theoretical models and methods that describe the nuclear many-body problem, which remains a critical challenge from the low-energy side~\cite{Bally:2022vgo}. The imaging power of high-energy nucleus collisions provides a complementary approach to the existing low-energy nuclear structure study~\cite{Giacalone:2023hwk}.

%\textcolor{red}{[Here, we need the text and equation for the W-S distribution for the NS]}\\
In the initial conditions of heavy-ion collisions, the nucleon positions are sampled from a modified Woods-Saxon distribution
\begin{align}
\rho^\mathrm{WS}(r) &= \frac{\rho_0}{1+\exp\left(\frac{r-R(\theta,\varphi)}{a_0}\right)}.
\label{eq:WS}
\end{align}
The deformation of the nuclear surface is described by
\begin{align}
R(\theta,\varphi) &= R_0\left(1+\beta_2\left[\cos\gamma Y_{20}(\theta,\varphi)+\sin\gamma Y_{22}(\theta,\varphi)\right]\right)
\label{eq:R}
\end{align}
where $R_0$ is the nuclear radius, $Y_{lm}$ are spherical harmonics, $\beta_2$ quantifies the strength of the quadrupole deformation, and the triaxial parameter $\gamma$ controls the relative order of the three axes in the intrinsic frame of the nuclei. The impacts of $\beta_2$ and $\gamma$ on the nuclear structure are illustrated in Fig.~\ref{fig:0}. Systematic studies of anisotropic flow using multi-particle azimuthal angle correlations show that several flow observables are sensitive to the nuclear quadrupole deformation $\beta_2$~\cite{Lu:2023fqd}. However, none of the flow observables (in terms of azimuthal angle correlations) have the power to distinguish the triaxial structures~\cite{Lu:2023fqd}, suggesting that different choices of observable may be needed to accomplish this, such as $v_n^{2}$-\mpt correlations~\cite{Giacalone:2019pca,Bally:2021qys,ALICE:2021gxt,Bally:2023dxi}, or multi-particle \pt correlations~\cite{Broniowski:2009fm,Giacalone:2020lbm}.

This paper will present a generic recursive algorithm for the genuine multi-particle correlations (cumulants) of \pt in Section~\ref{sec:algo}. The Monte Carlo models, introduced in Section~\ref{sec:MC}, will be used to examine the power of multi-particle cumulants of \pt, with the results presented in Section~\ref{sec:resu}. Finally, we will summarize in Section~\ref{sec:sum}.

\section{Algorithm for multi-particle \pt correlations}
\label{sec:algo}

%The multi-particle correlation has been widely used in high-energy nucleus collisions to investigate the initial conditionc~\cite{Bilandzic:2013kga,Moravcova:2020wnf}. It is ideal for constraining the quadrupole deformation $\beta_2$ and probing the triaxial structure $\gamma$~\cite{Jia:2021qyu,Jia:2022qgl,Lu:2023fqd}, which may further allow probing of the many-body interactions within the colliding nuclei. Among other choices of multi-particle correlations, the multi-particle \pt correlations introduced above give direct access to the initial state size/density. The previous study of the lower order \mpt fluctuations confirmed the power to probe the nuclear structure, i.e., nuclear deformation~\cite{Jia:2021qyu}. For the more complicated studies, such as fluctuations of nuclear structure, higher-order correlations involving more particles are necessary~\cite{Bhatta:2021qfk}. At the same time, direct implementations of these sophisticated multi-particle correlations with efficient calculations and accurate experimental corrections are not yet available. % I removed this paragraph from the introduction due to the repetition of the following paragraph -- YOU.

Multi-particle correlations have been a powerful tool in the study of heavy-ion collisions~\cite{Borghini:2000sa,Ollitrault:2023wjk}. One successful example is using multi-particle azimuthal angle correlations to study the final state anisotropic particle expansion, quantified by anisotropic flow $v_n$~\cite{Bilandzic:2013kga,Moravcova:2020wnf}. Systematic studies of anisotropic flow, flow fluctuations, and flow correlations show that the selected flow observables are sensitive to the nuclear quadrupole deformation $\beta_2$~\cite{Lu:2023fqd}. Still, none of the flow observables has the power to distinguish the triaxial structures~\cite{Lu:2023fqd}. For the study of multi-particle \pt correlations, the lower orders have been available and used in the investigations of the initial conditions in heavy-ion collisions and also explored for their potential in the study of nuclear structure~\cite{Jia:2021qyu,Bhatta:2021qfk}.
Meanwhile, it is challenging and computationally demanding for the higher orders when one manually derives the correlations. A generic recursive algorithm provides an ideal solution. It has been successfully developed for multi-particle azimuthal correlations~\cite{Moravcova:2020wnf} and used in the above-mentioned anisotropic flow studies~\cite{Lu:2023fqd}. Following a similar idea, we propose a generic recursive algorithm for multi-particle \mpt correlations in this paper.

In a single collision, many particles are produced. The averaged \pt correlation in this event is defined as
\begin{align}
\mptk{m} = \frac{\displaystyle\sum^M_{k_1\neq\ldots\neq k_m}w_{k_1}\cdot\ldots\cdot w_{k_m}p_{\mathrm{T},k_1}\cdot\ldots\cdot p_{\mathrm{T},k_m}}{\displaystyle\sum^M_{k_1\neq\ldots\neq k_m}w_{k_1}\cdot\ldots\cdot w_{k_m}},
\label{eq:ptcorr_general}
\end{align}
where $M$ is the number of particles within a certain kinematic range in this event and $w_{k_i}$ is a particle weight correcting for detector inefficiencies. Applying the constraint that $k_1\neq\ldots\neq k_m$ removes any auto-correlations such that the averaged \pt correlation measures the dynamic correlations due to event-by-event fluctuations and not the statistical correlations due to the finite number of tracks. The notation \mpt is used for the mean transverse momentum calculated event-by-event within a certain kinematic range and is distinct from the mean transverse momentum, $\la\pt\ra$, extrapolated to $\pt = 0$ from spectra analyses over a large event ensemble~\cite{ALICE:2010syw}. Similar to the Generic Framework for azimuthal angle correlations~\cite{Bilandzic:2013kga,Moravcova:2020wnf}, the numerator and denominator can be defined as
\begin{align}
N\la m\ra_{\pt} &= \displaystyle\sum^M_{k_1\neq\ldots\neq k_m}w_{k_1}\cdot\ldots\cdot w_{k_m}p_{\mathrm{T},k_1}\cdot\ldots\cdot p_{\mathrm{T},k_m},\label{eq:ptcorr_Num}\\
D\la m\ra_{\pt} &= \displaystyle\sum^M_{k_1\neq\ldots\neq k_m}w_{k_1}\cdot\ldots\cdot w_{k_m},\label{eq:ptcorr_Dn}
\end{align}
so that the averaged $m$-particle \pt correlation can be written as
\begin{align}
\mptk{m} = \frac{N\la m\ra_{\pt}}{D\la m\ra_{\pt}}.
\end{align}
The numerator and denominator in Eqs. \eqref{eq:ptcorr_Num} and \eqref{eq:ptcorr_Dn}, respectively, are calculated by constructing sums of different powers of the transverse momentum and particle weights
\begin{align}
P_{k}=\sum^M_{i=1} w_i^kp_{\mathrm{T},i}^k \quad, \quad W_{k}=\sum^M_{i=1} w_i^k.\label{ptcorr_vec}
\end{align}
Unlike the previously done analyses of azimuthal angle correlations \cite{Bilandzic:2013kga,Moravcova:2020wnf}, the first-order \pt correlation does not vanish but is simply \mpt. The first few transverse momentum correlations are given by
\begin{align}
\mptk{1} &= \frac{P_1}{W_1} = \frac{\sum_{i=1}^M w_ip_{\mathrm{T},i}}{\sum_{i=1}^M w_i} = \mpt,\label{eq:ptcorr_1pc}\\
\mptk{2} &= \frac{P_1^2-P_2}{W_1^2-W_2},\label{eq:ptcorr_2pc}\\
\mptk{3} &= \frac{P_1^3-3P_2P_1+2P_3}{W_1^3-3W_2W_1+2W_3}\label{eq:ptcorr_3pc}\\
\mptk{4} &= \frac{P_1^4-6P_2P_1^2+3P_2^2+8P_3P_1-6P_4}{W_1^4-6W_2W_1^2+3W_2^2+8W_3W_1-6W_4}\label{eq:ptcorr_4pc}
\end{align}
A general recursive formula can be constructed for Eqs. \eqref{eq:ptcorr_Num} and \eqref{eq:ptcorr_Dn}, allowing one to calculate a \pt correlation of any order from powers of $P_k$ and $W_k$
\begin{align}
N\la m\ra_{\pt} = \sum_{k=1}^m(-1)^{k-1}N\la m-k\ra_{\pt}\frac{(m-1)!}{(m-k)!}P_k\label{eq:ptcorr_num_rec}\\
D\la m\ra_{\pt} = \sum_{k=1}^m(-1)^{k-1}D\la m-k\ra_{\pt}\frac{(m-1)!}{(m-k)!}W_k\label{eq:ptcorr_den_rec}
\end{align}
where $N\la 0\ra_{\pt} \equiv 1$ and $D\la 0\ra_{\pt} \equiv 1$. Further written out applications of this recursive formula up to eighth order can be found in \ref{sec:appdyncorr}.

The event-averaged \pt correlations correspond to the raw sample moments of the \pt distribution within some kinematic region
\begin{align}
\la [\pt^{(m)}]\ra = \frac{\displaystyle\sum_\mathrm{events}W'_m[\pt^{(m)}]}{\displaystyle\sum_\mathrm{events}W'_m},
\label{eq:ptcorr_ev_avg}
\end{align}
where $W'_m$ is an event weight. The choice of this weight here is $W'_m = D\la m\ra_{\pt}$, which is the number of particle pairs similar to the weight used in anisotropic flow calculations to minimize the effects of multiplicity fluctuations. The higher-order moments contain contributions from the lower orders. The \pt cumulants, $\kappa_m$, (calculated from the correlations, $\la [\pt^{(m)}] \ra$) reveal to which degree there exists a genuine \pt correlation amongst many particles (m) which cannot be factorized into correlations amongst fewer particles. A recursive formula for the $m$th-order cumulant~\cite{Smith:2012cum} is given by
\begin{align}
\kappa_m = \la [\pt^{(m)}]\ra - \sum_{k=1}^{m-1}\begin{pmatrix} m-1\\k-1\end{pmatrix}\la [\pt^{(m-k)}]\ra\kappa_{k}.
\label{eq:ptcorr_cum}
\end{align}
One can expand this formula by grouping common terms to get a formula solely dependent on powers of $\la [\pt^{(m)}] \ra$. This has been done in \ref{sec:appdyncum} up to eighth order. These genuine correlations will be shown to be sensitive to the nuclear structure of the colliding nuclei.

%{\bf Suggestion: Emil copies and pastes what he had in the relevant section of his thesis (some higher orders should be provided either in the regular part or the appendix. Kris takes it over from there.}

\section{Introduction to the theoretical models}
\label{sec:MC}

%Brief introductions to the two MC models used in this study.

The Heavy Ion Jet INteraction Generator, HIJING~\cite{Wang:2000bf}, is a Monte Carlo event generator for parton and particle production in high-energy nuclear collisions. Based on QCD-inspired models for multiple jet production, it incorporates mechanisms such as multiple mini-jet production, soft excitation, nuclear shadowing of parton distribution functions, and jet interactions in dense hadronic matter. It has been reported that with the proper tuning, the HIJING model can describe particle production in high-energy proton-proton, proton-nucleus, and nucleus-nucleus collisions~\cite{Wang:2000bf}. Thus, the HIJING model is expected to provide reliable baseline predictions for the mean transverse momentum and its event-by-event fluctuations for heavy-ion collisions at the LHC.

Besides the HIJING model, A Multi-Phase Transport model, AMPT, is another popular Monte Carlo model that simulates nuclei collisions at relativistic energies~\cite{Lin:2004en}. The AMPT model run with String Melting consists of four main parts. First, the initial condition is based on the HIJING model~\cite{Wang:2000bf}, which describes how the nucleon distributions are arranged and moved inside the nuclei before they collide. Second, the Parton Cascade (ZPC)~\cite{Zhang:1997ej} simulates the partonic interactions inside the nuclei during the collision. Third, the hadronization, which is done via quark-coalescence~\cite{Chen:2005mr}, converts the partons into hadrons. Last, the hadronic interactions are modeled using the ART model~\cite{Li:1995pra}. The AMPT model can reproduce many experimental measurements from heavy-ion collisions, such as the production of various particle types and their collective behavior (the anisotropic flow). The  AMPT model with default settings has difficulty reproducing the centrality dependence of \mpt measurements in heavy-ion collisions at the LHC. Nevertheless, one can implement the nuclear structures in the initial state of the AMPT model. This allows one to examine if the imprint of the nuclear structure can be observed in the final state observables despite undergoing both partonic and hadronic interactions.

\begin{table}
\begin{center}
\begin{tabular}{c|c|c|c|c}
Data set \# 	&$\beta_2$	&$\gamma$	&$a_0$	&\# of events\\\hline\hline
1	&0	&0	&0.57 &$\sim$1M\\\hline
2	&0.18	&0	&0.57 &$\sim$1.5M\\\hline
3	&0.18	&27$^\circ$	&0.57 &$\sim$2.5M\\\hline
4	&0.18	&60$^\circ$	&0.57 &$\sim$2M\\\hline
\end{tabular}
\end{center}
\caption{Simulated \XeXe collisions data sets in AMPT with four different configurations of the $^{129}$Xe nuclear structure.}
\label{tab:ampt_prod}
\end{table}

%\textcolor{red}{\bf $\bullet$ We should also explain how we implement the different nuclear structures in the AMPT model. Emil, please add the table of AMPT parameters here.}\\

This paper implements four different settings for the $^{129}$Xe structure. The parameters relevant to the nuclear structure studies are listed in Table~\ref{tab:ampt_prod}, while other input parameters in the AMPT model are kept fixed between the data sets. Like many previous AMPT model studies~\cite{Lu:2023fqd}, the Lund string parameters are set to $a = 0.3$ and $b=0.15$, the screening mass to $\mu = 2.2814$. The time-step is 0.2 fm, and the number of time steps for the hadronic interactions within ART is set to 150, which gives a hadron cascade time of 30 fm/$c$.

%\textcolor{red}{Emil: Should a short discussion on the liquid-drop model and its assumptions be mentioned in this section? And in that case should it be renamed from MC models to just Models or something similar?}

\begin{table*}
\centering
\begin{tabularx}{\textwidth}{@{}c*1{>{\centering\arraybackslash}X}@{}*1{>{\centering\arraybackslash}X}@{}}
Final state	&Initial state	&Liquid-drop\\
cumulant    &cumulant   &model\\\hline
%% second-order cumulant
$\kappa_2$  &$\lla\left(\frac{\delta d_\perp}{d_\perp}\right)^2\rra$ &$\frac{1}{32\pi}\la \beta_2^2\ra$\\\hline
%% third-order cumulant
$\kappa_3$  &$\lla\left(\frac{\delta d_\perp}{d_\perp}\right)^3\rra$ &$\frac{\sqrt{5}}{896\pi^{3/2}}\la\cos(3\gamma)\beta_2^3\ra$\\\hline
%% fourth-order cumulant
$\kappa_4$  &$\left\langle\left(\frac{\delta d_\perp}{d_\perp}\right)^4\right\rangle-3\cdot\left\langle\left(\frac{\delta d_\perp}{d_\perp}\right)^2\right\rangle^2$ &$-\frac{3}{14336\pi^2}(7\la\beta_2^2\ra-5\la\beta_2^4\ra)$\\\hline
%% fifth-order cumulant
$\kappa_5$  &$\left\langle\left(\frac{\delta d_\perp}{d_\perp}\right)^5\right\rangle-10\cdot\left\langle\left(\frac{\delta d_\perp}{d_\perp}\right)^3\right\rangle\cdot\left\langle\left(\frac{\delta d_\perp}{d_\perp}\right)^2\right\rangle$ &$-\frac{5\sqrt{5}}{315392\pi^{5/2}}(11\la\cos(3\gamma)\beta_2^3\ra\la\beta_2^2\ra-5\la\beta_2^5\ra)$\\\hline
%% sixth-order cumulant
\multirow{2}[2]{*}{$\kappa_6$}  &$\left\langle\left(\frac{\delta d_\perp}{d_\perp}\right)^6\right\rangle-15\cdot\left\langle\left(\frac{\delta d_\perp}{d_\perp}\right)^4\right\rangle\cdot\left\langle\left(\frac{\delta d_\perp}{d_\perp}\right)^2\right\rangle$ &$\frac{5}{918412504\pi^{3}}(42042\la\beta_2^2\ra^3-5720\la\cos(3\gamma)\beta_2^3\ra^2$\\
&$+30\cdot\left\langle\left(\frac{\delta d_\perp}{d_\perp}\right)^2\right\rangle^3-10\cdot\left\langle\left(\frac{\delta d_\perp}{d_\perp}\right)^3\right\rangle^2$ &$-45045\la\beta_2^2\ra\la\beta_2^4\ra+8575\la\beta_2^6\ra+700\la\cos(6\gamma)\beta_2^6\ra)$\\\hline
%% seventh-order cumulant
\multirow{3}[5]{*}{$\kappa_7$}  &$\left\langle\left(\frac{\delta d_\perp}{d_\perp}\right)^7\right\rangle-21\cdot\left\langle\left(\frac{\delta d_\perp}{d_\perp}\right)^5\right\rangle\cdot\left\langle\left(\frac{\delta d_\perp}{d_\perp}\right)^2\right\rangle$    &$-\frac{15\sqrt{5}}{524812288}(2002\la\beta_2^2\ra^2\la\cos(3\gamma)\beta_2^3\ra$\\
&$+210\cdot\left\langle\left(\frac{\delta d_\perp}{d_\perp}\right)^3\right\rangle\cdot\left\langle\left(\frac{\delta d_\perp}{d_\perp}\right)^2\right\rangle^2$ &$+715\la\cos(3\gamma)\beta_2^3\ra\la\beta_2^4\ra$\\
&$-35\cdot\left\langle\left(\frac{\delta d_\perp}{d_\perp}\right)^3\right\rangle\cdot\left\langle\left(\frac{\delta d_\perp}{d_\perp}\right)^4\right\rangle$ &$+910\la\cos(3\gamma)\beta_2^5\ra\la\beta_2^2\ra-175\cos(3\gamma)\beta_2^7\ra)$\\\hline
%% eighth-order cumulant
\multirow{5}[14]{*}{$\kappa_8$}  &$\left\langle\left(\frac{\delta d_\perp}{d_\perp}\right)^8\right\rangle-28\cdot\left\langle\left(\frac{\delta d_\perp}{d_\perp}\right)^6\right\rangle\cdot\left\langle\left(\frac{\delta d_\perp}{d_\perp}\right)^2\right\rangle$  &$\frac{5}{142748942336\pi^4}(2144142\la\beta_2^2\ra^4-3063060\la\beta_2^2\ra^2\la\beta_2^4\ra$\\
&$+420\cdot\left\langle\left(\frac{\delta d_\perp}{d_\perp}\right)^4\right\rangle\left\langle\left(\frac{\delta d_\perp}{d_\perp}\right)^2\right\rangle^2$    &$-340\la\beta_2^2\ra\biggl(2288\la\cos(3\gamma)\beta_2^3\ra^2-35\bigl(49\la\beta_2^6\ra$\\
&$-35\left\langle\left(\frac{\delta d_\perp}{d_\perp}\right)^4\right\rangle^2-630\cdot\left\langle\left(\frac{\delta d_\perp}{d_\perp}\right)^2\right\rangle^4$ &$+4\la\cos(6\gamma)\beta_2^6\ra\bigr)\biggr)+25\biggl(21879\la\beta_2^4\ra^2$\\
&$+560\cdot\left\langle\left(\frac{\delta d_\perp}{d_\perp}\right)^3\right\rangle^2\cdot\left\langle\left(\frac{\delta d_\perp}{d_\perp}\right)^2\right\rangle$ &$+14144\la\cos(3\gamma)\beta_2^3\ra\la\cos(3\gamma)\beta_2^5\ra$\\
&$-56\cdot\left\langle\left(\frac{\delta d_\perp}{d_\perp}\right)^5\right\rangle\cdot\left\langle\left(\frac{\delta d_\perp}{d_\perp}\right)^3\right\rangle$   &$-35\bigl(79\la\beta_2^8\ra+16\la\cos(6\gamma)\beta_2^8\ra\bigr)\biggr)$\\\hline
\end{tabularx}
\caption{The cumulants of $d_\perp$ up to eighth order in a liquid-drop model potential averaged over random orientations. The first three entries are given in \cite{Jia:2021qyu}.}
\label{tab:liquiddrop}
\end{table*}

Following the idea established in \cite{Jia:2021qyu}, we consider the nuclei in a liquid-drop model with a sharp surface potential and in head-on collisions with a zero impact parameter. The mean transverse momentum is positively correlated with the inverse transverse size, $d_\perp$, in the overlapping region between the colliding nuclei
\begin{align}
    d_\perp = \sqrt{N_\mathrm{part}/\la r_\perp^2\ra},
    \label{eq:dperp}
\end{align}
where $N_\mathrm{part}$ is the number of nucleons participating in the collision and $r_\perp$ is the transverse radius. Within the liquid-drop model, the dependence of the central moment of $d_\perp$ on the nuclear structure parameters can be roughly approximated with the following relation \cite{Jia:2021qyu}
\begin{align}
    \frac{\delta d_\perp}{d_\perp}&=\sqrt{\frac{5}{16\pi}}\beta_2\biggl(\cos(\gamma)D_{0,0}^2(\Omega)    \label{eq:IScumulant}\\
    &+\frac{\sin(\gamma)}{\sqrt{2}}[D_{0,2}^2(\Omega)+D_{0,-2}^2(\Omega)]\biggr),\nonumber
\end{align}
where $D_{m,m'}^2$ are the elements of the Wigner D-matrix and $\Omega$ represents the Euler angles. The expressions for the higher-order central moments are obtained by integrating over the Euler angles
\begin{align}
    \biggl\la\biggl(\frac{\delta d_\perp}{d_\perp}\biggr)^n\biggr\ra &= \beta_2^n\biggl(\frac{5}{16\pi}\biggr)^{n/2}\int \biggl(\cos(\gamma)D_{0,0}^2\label{eq:integral}\\
    &+\frac{\sin(\gamma)}{\sqrt{2}}[D_{0,2}^2+D_{0,-2}^2]\biggr)^n \frac{d\Omega}{8\pi^2}\nonumber
\end{align}
from which the cumulants can be constructed, and the final cumulants of $n$-th order are scaled by $1/2^{n-1}$ to account for the independent orientations of the two nuclei. Table \ref{tab:liquiddrop} shows the liquid-drop model estimates for the higher-order cumulants.

\section{Results}
\label{sec:resu}
\begin{figure}[t]
  \includegraphics[width=\linewidth]{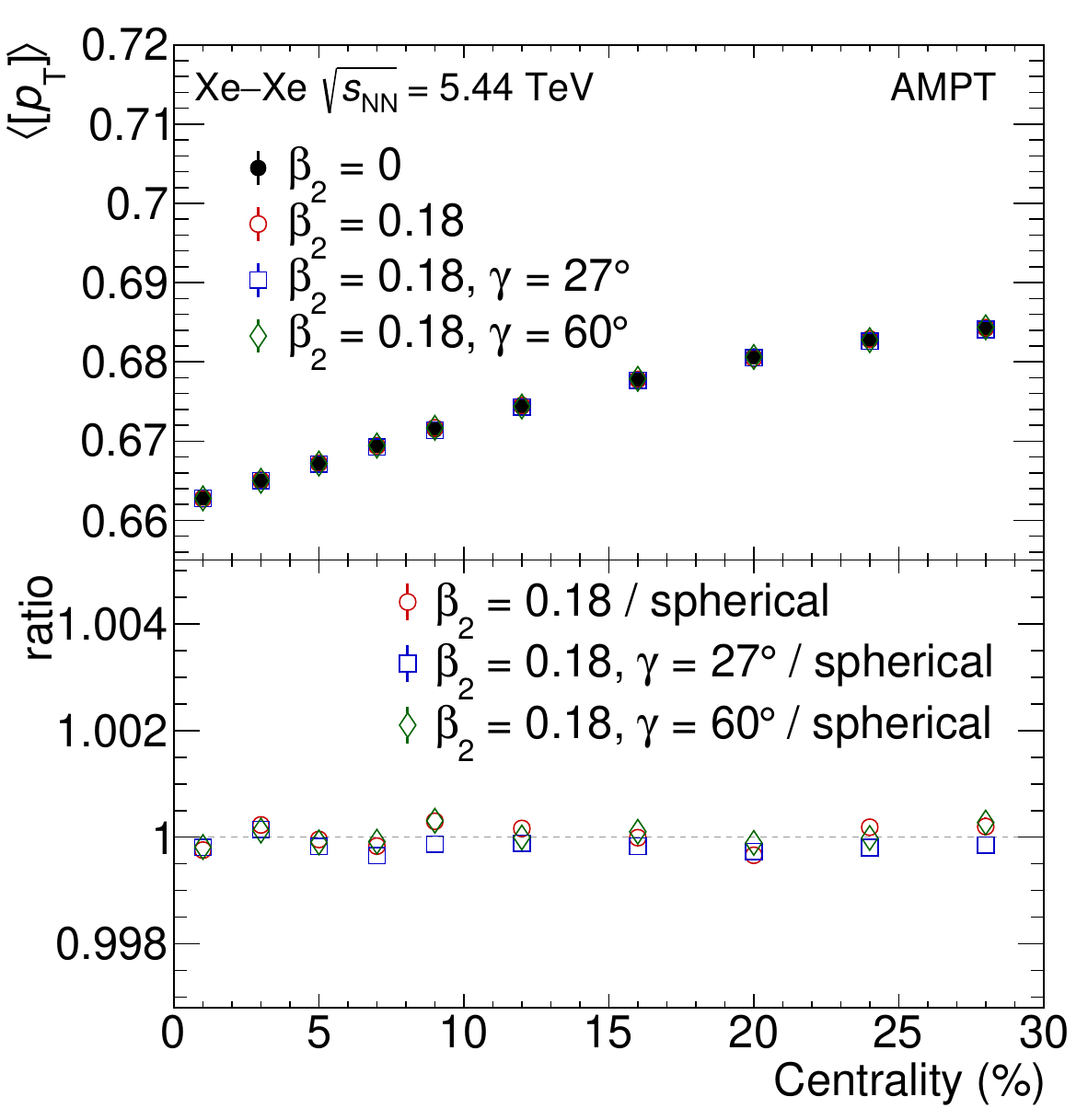}
\caption{The mean transverse momentum, \mmpt, in \XeXe collisions at \snn = 5.44 TeV as a function of centrality for four configurations of the nuclear structure simulated with the AMPT model. }
\label{fig:1}       
\end{figure}
\begin{figure}[t]
  \includegraphics[width=\linewidth]{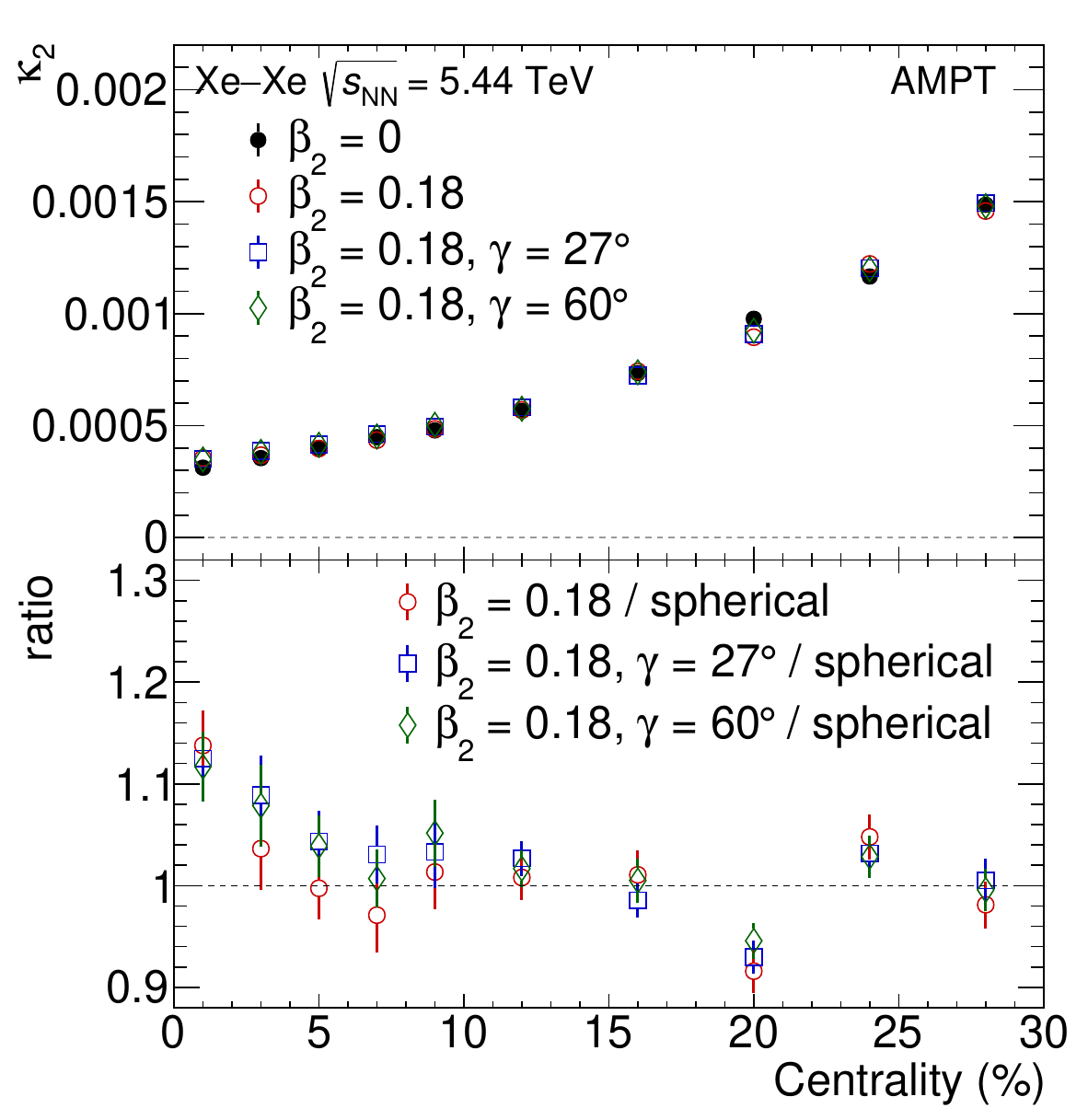}
\caption{The second-order mean transverse momentum cumulant, $\kappa_2$, in \XeXe collisions at \snn = 5.44 TeV as a function of centrality for four configurations of the nuclear structure simulated with the AMPT model.}
\label{fig:2}      
\end{figure}
\begin{figure}[t]
  \includegraphics[width=\linewidth]{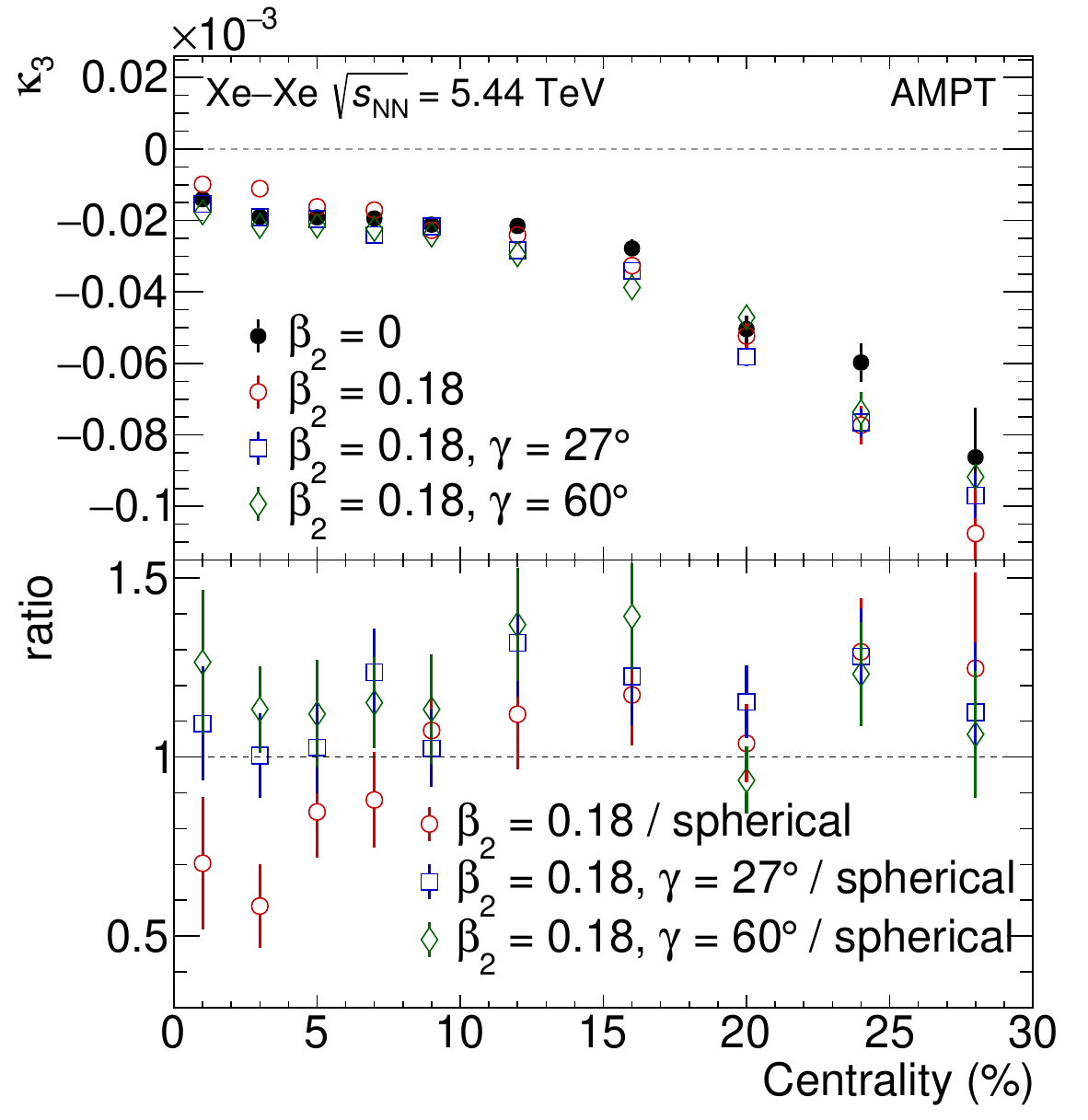}
\caption{The third-order mean transverse momentum cumulant, $\kappa_3$, in \XeXe collisions at \snn = 5.44 TeV as a function of centrality for four configurations of the nuclear structure simulated with the AMPT model.}
\label{fig:3}
\end{figure}
\begin{figure}[t]
  \includegraphics[width=\linewidth]{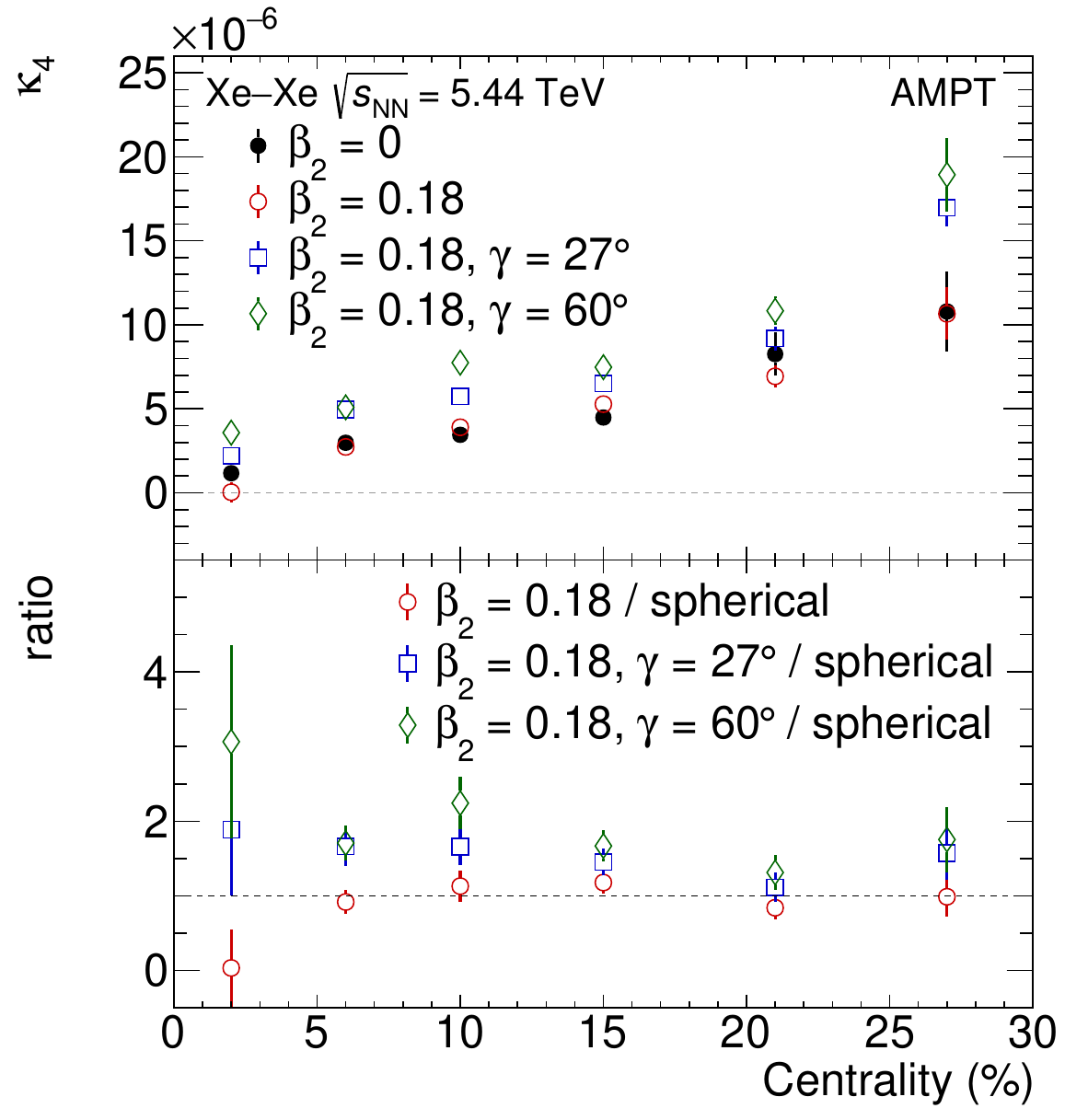}
\caption{The fourth-order mean transverse momentum cumulant, $\kappa_4$, in \XeXe collisions at \snn = 5.44 TeV as a function of centrality for four configurations of the nuclear structure simulated with the AMPT model.}
\label{fig:4}
\end{figure}
%

%here we need to add a table of derived sensitivities of higher order [pt] fluctuations/cumulants to the deformation parameters, based on the liquid drop model, in addition to what is listed in Jiangyong's paper (which has the first three orders).

%{\bf $\bullet$ AMPT results on the first few orders with different nuclear structures}

%\textcolor{red}{A general question: do we want first to discuss AMPT results of every cumulant and then liquid-drop model calculations, or discuss cumulant order by order with both AMPT and liquid-drop model.}

Figure \ref{fig:1} shows the \mmpt in \XeXe collisions at \snn = 5.44 TeV as a function of centrality. The top panel shows the \mmpt for a spherical, prolate, triaxial, and oblate configuration of \Xe, and the bottom panel shows the ratio of \mmpt in collisions of deformed nuclei against the \mmpt calculated in collisions of spherical nuclei. The different ratios are consistent with unity suggesting that to first order, the transverse momentum cumulants are not a useful probe of the nuclear structure as the \mmpt does not differ between the spherical ($\beta_2=0$), prolate ($\beta_2=0.18$), triaxial ($\beta_2=0.18$, $\gamma=27^\circ$), and oblate ($\beta_2=0.18$, $\gamma=60^\circ$) nuclei. This is consistent with the expectation from the liquid-drop model; the integral in \eqref{eq:integral} yields zero when evaluated for $n=1$. 

The centrality dependence of \mpt fluctuations, in the form of the second-order cumulant or variance, $\kappa_2$, is shown in Fig. \ref{fig:2}. The $\kappa_2$ result increases by about 10-15\% in the central collisions for $\beta_2 \neq 0$ compared to the $\kappa_2$ for spherical nuclei. However, the results show no distinction between $\kappa_2$ calculations from prolate, triaxial, and oblate nuclei. This agrees with the expectation from the liquid-drop model, which predicts a positive contribution (an increase) in the second-order cumulant that is proportional to $\la\beta_2^2\ra$ and no sensitivity to the $\gamma$ parameter, shown in Table~\ref{tab:liquiddrop}. 
Therefore, we recommend that $\kappa_2$ should be considered as an additional probe to the anisotropic flow measurements~\cite{ALICE:2018lao} for constraining the parameter of \Xe. In particular, it should be included in the Bayesian fits on the experimental measurements to extract reliable deformation parameters from the high-energy heavy-ion collisions.

%\textcolor{red}{We might consider commenting that the difference in beta2 is significant and well beyond the statistical and systematical uncertainties in experiments. Thus, one should be able to make the data and model comparisons to constrain the beta2 parameter or consider it in the global Bayesian analysis to extract precise beta2 from high-energy EXP.}

%\textcolor{red}{We should add the discussions that this is expected from the liquid-drop model perspective, as a non-zero $\beta_2$ will boost the kappa2 results, while it is expected to be insensitive to $\gamma$.}

The centrality dependence of the third-order cumulant, $\kappa_3$, is shown in Figure~\ref{fig:3}, where three distinct trends in central collisions, depending on the triaxiality parameter $\gamma$, are observed. Compared to the calculations from spherical nuclei, the magnitude of $\kappa_3$ is around 30\% larger for oblate nuclei ($\gamma = 60^\circ$) and around 30\% smaller for prolate nuclei ($\gamma = 0^\circ$) in central collisions albeit with large statistical uncertainties. The triaxial \Xe ($\gamma = 27^\circ$) leads to a value of $\kappa_3$ consistent with the spherical case ($\beta_2=0$) in the 10\% most central collisions. The emergence of the sensitivity to the triaxiality in $\kappa_3$ is unsurprising as triaxiality is a three-point structure, so a three-particle correlation is needed to measure it. This also explains why $\kappa_2$ can distinguish the overall quadrupole deformation ($\beta_2=0.18$) from the spherical case but offers no additional information: the two-particle $\kappa_2$ can only probe two axes of the nuclei at a time. In addition, the liquid-drop model predicts a positive contribution from the $\gamma$ parameter in the prolate case ($\cos(3\gamma) = 1$), zero contribution from fully triaxial ($\cos(3\gamma)=0$), and negative contribution from the oblate case ($\cos(3\gamma)=-1$). The AMPT calculations follow the same trend. However, as the $\kappa_3$ values are negative, the ordering of the points with different $\gamma$ parameters is inverted in the ratio plot, shown in the bottom panel of Fig.~\ref{fig:3}. The above results further underline the importance of $\kappa_3$ in exploring triaxial nuclear structure. It suggests that not only the correlations between \pt and $v_2^{2}$, which have been widely used~\cite{Bozek:2016yoj,Giacalone:2020dln,ALICE:2021gxt} but also the third order \mpt fluctuations in central collisions should be considered in the comparisons between model calculations and experimental measurements from the LHC experiments.
 
 %\textcolor{red}{We should also add the discussions that this is expected from the liquid-drop model perspective, as $\gamma = 0$ will increase the kappa3 results (thus, red markers are higher and green are the smallest, while $\gamma = 27$ is roughly compatible with $\beta_2 = 0$ because $\cos (3 \gamma) = 0$. Here, we might also note that the ordering of various kappa3 results is inverse because they are divided by negative values.} 

Figure \ref{fig:4} shows the centrality dependence of the fourth-order cumulant, $\kappa_4$, and the comparison between the different nuclear structures. Similar to $\kappa_3$, $\kappa_4$ results can distinguish between different triaxial shapes in central collisions despite sizable uncertainties. The collisions of oblate nuclei yield the largest $\kappa_4$ in central collisions. The $\kappa_4$ for the triaxial case is smaller than the oblate case but larger than for spherical nuclei. The prolate nuclei yield the smallest $\kappa_4$ in central collisions, almost a 100\% difference with respect to the spherical case. This does not match the prediction from the simple liquid-drop model, where the fourth-order cumulant is expected to be proportional to some combination of the moments of $\beta_2$ only. However, a dependence of the initial state cumulant $(\delta d_\perp/d_\perp)^4-3(\delta d_\perp/d_\perp)^2$ on the $\gamma$ parameter is also seen in MC-Glauber simulations of various deformed collision systems \cite{Jia:2021qyu}. The ordering of the calculations in the bottom plot of Fig. \ref{fig:4} is similar to that in Fig. \ref{fig:3} (bottom), but in this case, the ordering is not inverted, suggesting that $\gamma$ either has an opposite contribution to $\kappa_4$ compared to $\kappa_3$ or that the $\gamma$ affects the magnitude of the cumulants. That $\kappa_4$ is sensitive to the $\gamma$ parameter is surprising based on the leading-order approximation of the liquid-drop model, and it would be interesting to see whether $\kappa_4$ exhibits similar sensitivity within hydrodynamic models. 
%{\bf $\bullet$ HIJING results on low to high orders.}

\begin{figure*}[t]
  \includegraphics[width=\linewidth]{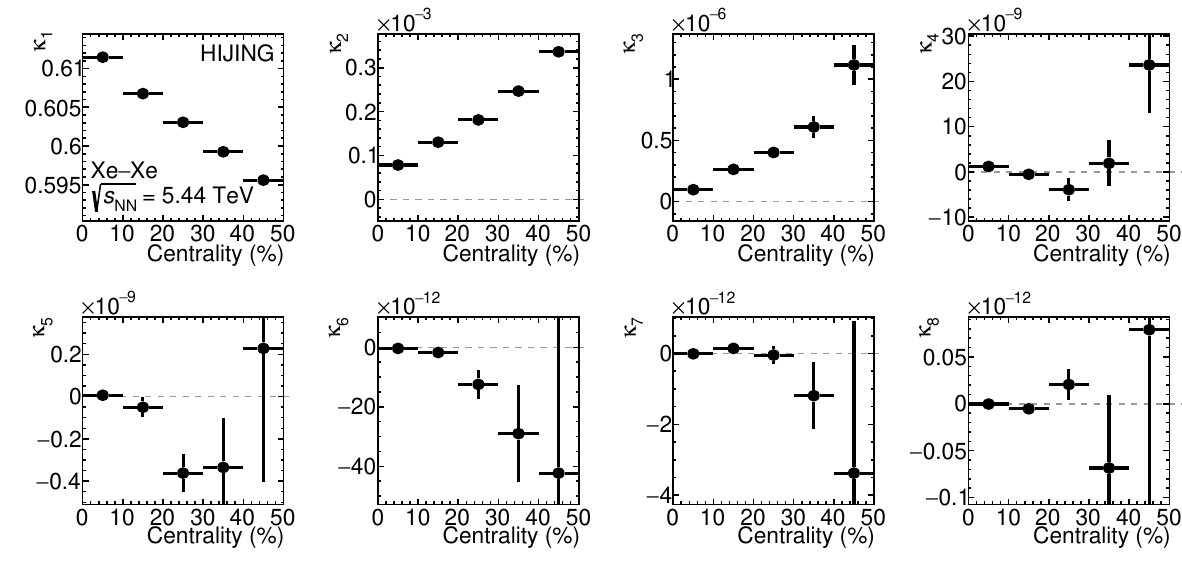}
% figure caption is below the figure
\caption{The cumulants of the mean transverse momentum, $\kappa_n$ ($n\leq 8$), in \XeXe collisions at \snn = 5.44 TeV as a function of centrality simulated with the HIJING model.}
\label{fig:5}       % Give a unique label
\end{figure*}
A baseline prediction for the various orders of $\kappa_n$ $(n\leq 8)$ is given with the HIJING model in Fig. \ref{fig:5}. Within the HIJING model, the fluctuations of the transverse momentum follow that of a superposition of independent sources, which can be both short- and long-range in nature \cite{Bhatta:2021qfk,Wang:1991hta}. The first three orders of cumulants, $\kappa_{1}$-$\kappa_3$, are predicted to be strictly positive across the centrality range. $\kappa_4$ is positive in central collisions, changes signs in semi-central collisions and then becomes positive again above 30\% centrality. The higher-order cumulants ($n>4$) fluctuate around zero in central and semi-central collisions, except $\kappa_6$, which is strictly negative. The non-zero higher-order cumulants suggest that the fluctuations of the mean transverse momentum are not trivial even in heavy-ion collisions viewed as a superposition of \pp collisions. 

\section{Summary}
\label{sec:sum}

Multi-particle correlations and cumulants of transverse momentum constrain the size and size fluctuations in the initial conditions of heavy-ion collisions, making them an ideal probe of the nuclear structure of colliding nuclei at ultra-relativistic colliders. This paper proposes a new generic algorithm of multi-particle \pt correlations that enables an efficient and precise study of \pt fluctuations up to arbitrary orders. Using the AMPT transport model with various settings of nuclear structure, including spherical, prolate, triaxial, and oblate shapes of \Xe, we investigate the impact of nuclear structure on the first four orders of \mpt fluctuations. We find that at first order, the \pt cumulants are insensitive to the nuclear structure and that at least second order is required to probe the deformation of the colliding nuclei. The variance, $\kappa_2$, shows a $\beta_2$-dependent increase in central collisions but cannot distinguish between the prolate, triaxial, or oblate cases consistent with the expectation from the simple liquid-drop model. The third- and fourth-order cumulants, $\kappa_3$ and $\kappa_4$, show a $\gamma$-dependent splitting of the calculated values in the most central collisions with opposite contribution. The $\gamma$-dependence of $\kappa_4$ is not expected from the liquid-drop model prediction, suggesting that a more complex model of the interplay between nuclear structure and size fluctuations is necessary to understand this effect.

These studies show the impact of nuclear structure on the \mpt fluctuations and how they can be utilized to probe the shape of the colliding nuclei at the energy frontier (at the level of TeV). The \pt cumulants are, through their relation to the size fluctuations, coupled to the moments of the deformation parameters. Further study of these relations and measurements of the cumulants at experiments such as those at the LHC can introduce a new avenue for studying nuclear structure in high-energy heavy-ion collisions, which is complementary to existing low-energy measurements and can serve as experimental validation of low-energy theoretical predictions. Furthermore, including the \pt cumulants as experimental inputs in Bayesian analyses can provide independent constraints on the deformation parameters. 

The study of nuclear structure with heavy-ion collisions is an evolving field, and the presented method and observables can provide a useful approach in potential future runs of deformed nuclei species at the LHC to constrain the shape of nuclei across the nuclide chart.

%End of our recommendation on nuclear structure at the energy frontier. 

\begin{acknowledgements}
We thank Chunjian Zhang for his help on the AMPT code and Anna Ingmer Boye for her initial contributions to the work. The authors are funded by the European Union (ERC, InitialConditions), VILLUM FONDEN (grant number 00025462), and Danmarks Frie Forskningsfond (Independent Research Fund Denmark). 
\end{acknowledgements}

{\bf Data availability:} The manuscript has associated data in a data repository [Authors’ comment: Data are available upon request.]

% BibTeX users please use one of
%\bibliographystyle{spbasic}      % basic style, author-year citations
%\bibliographystyle{spmpsci}      % mathematics and physical sciences
\bibliographystyle{utphys}       % APS-like style for physics
\bibliography{bibliography.bib}   % name your BibTeX data base

\appendix
\section{Multi-particle \pt correlations}
\label{sec:appdyncorr}
\begin{align}
\mptk{1} &= \frac{P_1}{W_1}
\end{align}
\begin{align}
\mptk{2} &= \frac{P_1^2-P_2}{W_1^2-W_2}
\end{align}
\begin{align}
\mptk{3} &= \frac{P_1^3-3P_2P_1+2P_3}{W_1^3-3W_2W_1+2W_3}
\end{align}
\begin{align}
\mptk{4} &= \frac{P_1^4-6P_2P_1^2+3P_2^2+8P_3P_1-6P_4}{W_1^4-6W_2W_1^2+3W_2^2+8W_3W_1-6W_4}
\end{align}
\begin{align}
\mptk{5} &= \left[P_1^5-10P_2P_1^3+15P_2^2P_1\right.\\
&\left.+20P_3P_1^2-20P_3P_2-30P_4P_1+24P_5\right]\nonumber\\
&/\left[W_1^5-10W_2W_1^3+15W_2^2W_1+20W_3W_1^2\right.\nonumber\\
&\left.-20W_3W_2-30W_4W_1+24W_5\right]\nonumber
\end{align}
\begin{align}
\mptk{6} &= \left[P_1^6-15P_2P_1^4+45P_1^2P_2^2\right.\\
&-15P_2^3+40P_3P_1^3-120P_3P_2P_1\nonumber\\
&+40P_3^2-90P_4P_1^2+90P_4P_2\nonumber\\
&\left.+144P_5P_1-120P_6\right]/\left[W_1^6-\right.\nonumber\\
&15W_2W_1^4+45W_1^2W_2^2-15W_2^3\nonumber\\
&+40W_3W_1^3-120W_3W_2W_1\nonumber\\
&+40W_3^2-90W_4W_1^2+90W_4W_2\nonumber\\
&\left.+144W_5W_1-120W_6\right]\nonumber
\end{align}
\begin{align}
\mptk{7} &= \left[P_1^7-21P_2P_1^5+105P_1^3P_2^2\right.\\
&-105P_2^3P_1+70P_3P_1^4-420P_3P_2P_1^2\nonumber\\
&+210P_3P_2^2+280P_3^2P_1-210P_4P_1^3\nonumber\\
&+630P_4P_2P_1-420P_4P_3+504P_5P_1^2\nonumber\\
&\left.-504P_5P_2-840P_6P_1+720P_7\right]/\left[W_1^7\right.\nonumber\\
&-21W_2W_1^5+105W_1^3W_2^2\nonumber\\
&-105W_2^3W_1+70W_3W_1^4-420W_3W_2W_1^2\nonumber\\
&+210W_3W_2^2+280W_3^2W_1-210W_4W_1^3\nonumber\\
&+630W_4W_2W_1-420W_4W_3+504W_5W_1^2\nonumber\\
&\left.-504W_5W_2-840W_6W_1+720W_7\right]\nonumber
\end{align}
\begin{align}
\mptk{8} &= \left[P_1^8-28P_2P_1^6+210P_2^2P_1^4\right.\\
&-420P_2^3P_1^2+105P_2^4+112P_3P_1^5\nonumber\\
&-1120P_3P_2P_1^3+1680P_3P_2^2P_1+1120P_3^2P_1^2\nonumber\\
&-1120P_3^2P_2-420P_4P_1^4+2520P_4P_2P_1^2\nonumber\\
&-1260P_4P_2^2-3360P_4P_3P_1+1260P_4^2\nonumber\\
&+1344P_5P_1^3-4032P_5P_2P_1+2688P_5P_3\nonumber\\
&-3360P_6P_1^2+3360P_6P_2+5760P_7P_1\nonumber\\
&\left.-5040P_8\right]/\left[W_1^8-28W_2W_1^6+210W_2^2W_1^4\right.\nonumber\\
&-420W_2^3W_1^2+105W_2^4+112W_3W_1^5\nonumber\\
&-1120W_3W_2W_1^3+1680W_3W_2^2W_1+1120W_3^2W_1^2\nonumber\\
&-1120W_3^2W_2-420W_4W_1^4+2520W_4W_2W_1^2\nonumber\\
&-1260W_4W_2^2-3360W_4W_3W_1+1260W_4^2\nonumber\\
&+1344W_5W_1^3-4032W_5W_2W_1+2688W_5W_3\nonumber\\
&-3360W_6W_1^2+3360W_6W_2+5760W_7W_1\nonumber\\
&\left.-5040W_8\right]\nonumber
\end{align}

\section{Multi-particle \pt cumulants}
\label{sec:appdyncum}
\begin{align}
\kappa_1 &= \mmpt\\    
\kappa_2 &= \mmptk{2} - \mmpt^2\\    
\kappa_3 &= \mmptk{3}-3\mmptk{2}\mmpt + 2\mmpt^3\\
\kappa_4 &= \mmptk{4}-4\mmptk{3}\mmpt-3\mmptk{2}^2\\
&+12\mmptk{2}\mmpt^2-6\mmpt^4\nonumber\\
\kappa_5 &= \mmptk{5}-5\mmptk{4}\mmpt-10\mmptk{3}\mmptk{2}\\
&+30\mmptk{2}^2\mmpt+20\mmptk{3}\mmpt^2\nonumber\\
&-60\mmptk{2}\mmpt^3+24\mmpt^5\nonumber\nonumber\\
\kappa_6 &= \mmptk{6}-6\mmptk{5}\mmpt-15\mmptk{4}\mmptk{2}\\
&-10\mmptk{3}^2+30\mmptk{2}^3+30\mmptk{4}\mmpt^2\nonumber\\
&+120\mmptk{3}\mmptk{2}\mmpt-270\mmptk{2}^2\mmpt^2\nonumber\\
&-120\mmptk{3}\mmpt^3+360\mmptk{2}\mmpt^4\nonumber\\
&-120\mmpt^6\nonumber\\
\kappa_7 &= \mmptk{7}-7\mmptk{6}\mmpt -21\mmptk{5}\mmptk{2}\\
&+42\mmptk{5}\mmpt^2-35\mmptk{4}\mmptk{3}\nonumber\\
&+210\mmptk{4}\mmptk{2}\mmpt-210\mmptk{4}\mmpt^3\nonumber\\
&+140\mmptk{3}^2\mmpt+210\mmptk{3}\mmptk{2}^2\nonumber\\
&-1260\mmptk{3}\mmptk{2}\mmpt^2+840\mmptk{3}\mmpt^4\nonumber\\
&-630\mmptk{2}^3\mmpt+2520\mmptk{2}^2\mmpt^3\nonumber\\
&-2520\mmptk{2}\mmpt^5+720\mmpt^7\nonumber\\
\kappa_8 &= \mmptk{8}-8\mmptk{7}\mmpt\\
&-28\mmptk{6}\mmptk{2}+56\mmptk{6}\mmpt^2\nonumber\\
&-56\mmptk{5}\mmptk{3}+336\mmptk{5}\mmptk{2}\mmpt\nonumber\\
&-336\mmptk{5}\mmpt^3-35\mmptk{4}^2\nonumber\\
&+560\mmptk{4}\mmptk{3}\mmpt+420\mmptk{2}^2\mmptk{4}\nonumber\\
&-2520\mmptk{4}\mmptk{2}\mmpt^2+1680\mmptk{4}\mmpt^4\nonumber\\
&+560\mmptk{3}^2\mmptk{2}-1680\mmptk{3}^2\mmpt^2\nonumber\\
&-5040\mmptk{3}\mmptk{2}^2\mmpt+13440\mmptk{3}\mmptk{2}\mmpt^3\nonumber\\
&-6720\mmptk{3}\mmpt^5-630\mmptk{2}^4+10080\mmptk{2}^3\mmpt^2\nonumber\\
&-25200\mmptk{2}^2\mmpt^4+20160\mmptk{2}\mmpt^6-5040\mmpt^8\nonumber
\end{align}

\end{document}